\documentclass[review, 12pt]{elsarticle}
\usepackage[utf8]{inputenc}
\usepackage{amsmath}
\usepackage[plainpages=false,pdfpagelabels,unicode]{hyperref}
\usepackage[pdftex]{xcolor}
\usepackage[margin=2cm]{geometry}
\biboptions{numbers,sort&compress}
\usepackage{soul}
\usepackage{cancel}

\newcommand{\TiON}[0]{Ti$_{1-\delta}$O$_x$N$_{1-x}$}
\newcommand{\TiNx}[0]{TiN$_{x}$}
\newcommand{\TiNy}[1]{TiN$_{{#1}}$}

\newcommand{\NV}[1]{\mathrm{N}^{{#1}\mathrm{V}_\mathrm{Ti}}}
\newcommand{\TiNone}[0]{{\TiNy{1.37}}}
\newcommand{\TiNtwo}[0]{\TiNy{1.35}}
\newcommand{\TiNthree}[0]{\TiNy{1.23}}
\newcommand{\TiNfour}[0]{\TiNy{1.18}}

\begin{document}

\author[DPPT]{Pavel Ondračka}
\author[MCh]{Pauline Kümmerl}
\author[MCh]{Marcus Hans}
\author[MCh]{Stanislav Mráz}
\author[DPA]{Daniel Primetzhofer}
\author[MUL]{David Holec}
\author[DPPT]{Petr Vašina}
\author[MCh]{Jochen M. Schneider}

\address[DPPT]{Department of Plasma Physics and Technology, Faculty of Science, Masaryk University, Kotlářská 2, CZ-61137 Brno, Czech Republic}
\address[MCh]{Materials Chemistry, RWTH Aachen University, Kopernikusstr. 10, D-52074 Aachen, Germany}
\address[DPA]{Department of Physics and Astronomy, Uppsala University, L\"{a}gerhyddsv\"{a}gen 1, S-75120 Uppsala, Sweden}
\address[MUL]{Department of Materials Science, Montanuniversität Leoben, Franz-Josef-Strasse 18, A-8700 Leoben, Austria}

\title{Prediction and identification of point defect fingerprints in X-ray photoelectron spectra of TiN$_x$ with 1.18 $\le x \le$ 1.37}

\date{\today}

\begin{graphicalabstract}
   \includegraphics[width=\linewidth]{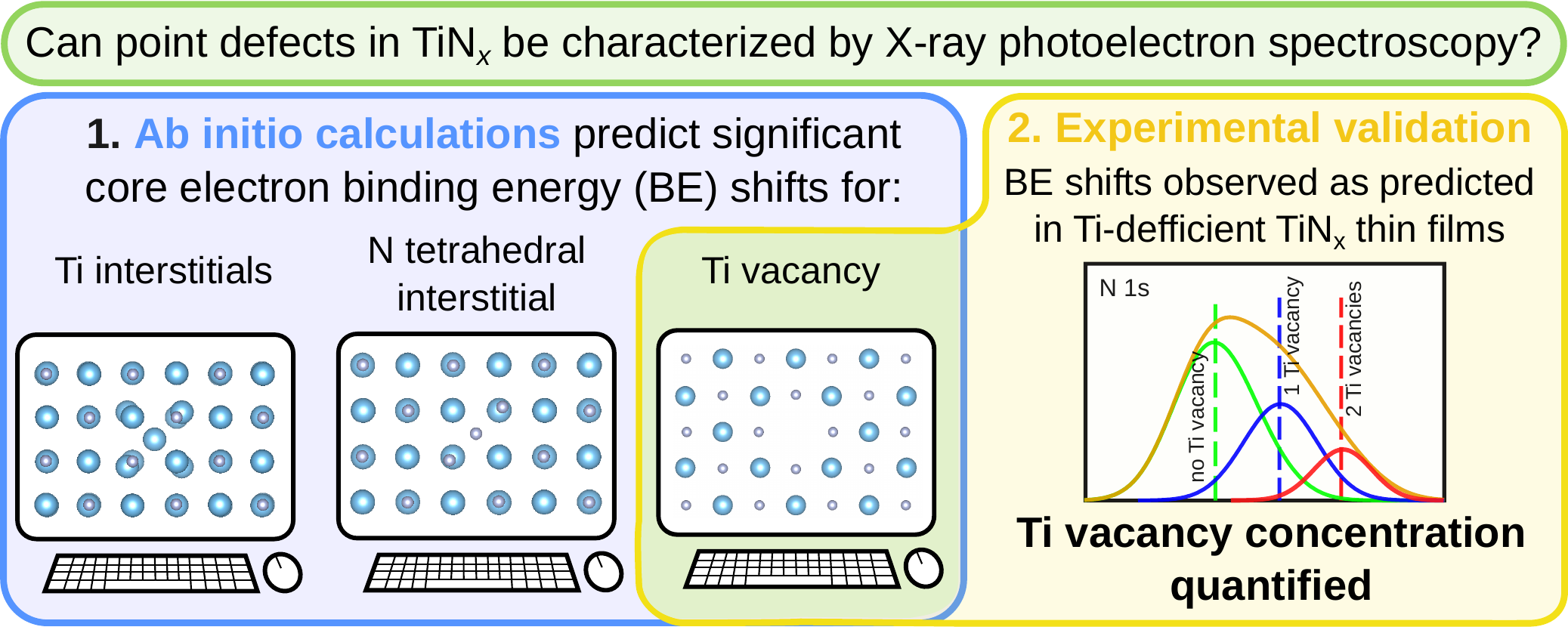}
\end{graphicalabstract}

\begin{abstract}
We investigate the effect of selected N and Ti point defects in $B$1 TiN on N\,1s and Ti\,2p$_{3/2}$  binding energies (BE) by experiments and {\it ab initio} calculations.
X-ray photoelectron spectroscopy (XPS) measurements of \TiNx{} films with 1.18 $\le x \le$ 1.37
reveal additional N\,1s spectral components at lower binding energies.
{\it Ab initio} calculations predict that these components are caused by either
Ti vacancies, which induce a N\,1s BE shift of $-0.54$\,eV in its first N neighbors,
and/or N tetrahedral interstitials,
which have their N\,1s BE shifted by $-1.18$\,eV and
shift the BE of their first N neighbors by $-0.53$\,eV.
However, based on {\it ab initio} data the tetrahedral
N interstitial is estimated to be unstable at room temperature.
We, therefore, unambiguously attribute the N\,1s spectral components at lower BE in Ti-deficient \TiNx{} thin films to the presence of Ti vacancies.
Furthermore, it is demonstrated that the vacancy concentration in Al-capped Ti-deficient TiN$_x$
can be quantified with the here proposed correlative method based on measured and predicted BE data.
Our work highlights the potential of {\it ab initio}-guided XPS measurements
for detecting and quantifying point defects in $B$1 \TiNx{}.
\end{abstract}


\maketitle

\section{Introduction}

$B$1 titanium nitride (TiN) is a well known material with applications like orthopedic implants~\cite{vanHove2015}, wear protection \cite{Hedenqvist1990} or decorative coatings~\cite{Nose2001} due to its excellent properties: high hardness, wear resistance~\cite{Santecchia2015} and thermal stability~\cite{Hultman2000}. In recent years, novel electrical applications such as solar cell contacts~\cite{Lu2020, Yang2019} or electrodes~\cite{Li2024} have been explored, since TiN possesses high electrical conductivity. Furthermore, TiN has also attracted some interest for plasmonics applications~\cite{Chang2019, Catellani2020, Mahajan2024} offering an alternative to traditional metals like gold or silver~\cite{Naik2013}. 
Properties of \TiNx{} are significantly influenced by the stoichiometry which is mainly facilitated by point defects: most notably N vacancies for $x < 1$, and Ti vacancies or N interstitials for $x > 1$~\cite{Mirguet2006, Schmid1998, Smith1999}.
Specifically, the presence of both N and Ti vacancies was shown to reduce elastic modulus~\cite{Zhang2021}, N vacancies were also shown to reduce the bulk modulus~\cite{Guemmaz2001} and increase hardness~\cite{Lee2005,Shin2003}. Importantly, the \TiNx{} optical performance can be tuned by a stoichiometry changes through both Ti and N vacancy introduction such that it can exhibit plasmonic properties from near infrared to long-wavelength infrared range~\cite{Catellani2020, Judek2021}. Point defects also play an important role during O incorporation~\cite{Conntable2024}, in fact Ti vacancies are created during the early stages of oxidation~\cite{Zimmermann2009, Ponon2015}. Therefore, knowledge of the specific point defect state is crucial in understanding the structure–property relationship in (not only) TiN thin films.

Several methods have been used in the past to characterize point defects in TiN, like electron energy loss spectroscopy~\cite{Zhang2021, Mirguet2006}, Raman spectroscopy~\cite{Ponon2015}, positron annihilation spectrocopy~\cite{Abiyev2024} or M\"{o}ssbauer spectroscopy~\cite{Qi2019}. Reliable quantification of point defects TiN, as well as in other derived cubic nitrides, however, still remains a challenge and novel methods are being developed. It was recently shown for WN$_{0.5}$ that the N vacancies can be imaged at the atomic scale by HRTEM when the vacancy concentrations is high~\cite{Chen2023}, or a method based on atom probe tomography has been proposed to quantify both metal and nitrogen vacancies in TiAlN nanolamella coatings~\cite{Qiu2021}, however this approach and its accuracy has been questioned~\cite{Hans2023}. 

Our recent work showed that it is possible to detect and even quantify the concentration of Ti vacancies in \TiON{} thin films using X-ray photoelectron spectroscopy (XPS)~\cite{Ondracka2022}. However, only the effect of Ti vacancies was studied in detail. The impact of other possible point defects---such as interstitials, whose presence was also suggested by the XRD results---on the XPS spectra was not thoroughly investigated.
This vacancy quantification work was enabled by guiding the XPS analysis with {\it ab initio} calculations that allow to precisely predict
core electron binding energies (BE) shifts using the density functional theory together with core-hole approaches~\cite{Ozaki2017, Aarva2019, Kahk2022}.
Such a combined approach can help to
explain spectral features observed experimentally~\cite{Panepinto2020}
and to obtain important information about the structure of the materials~\cite{Ondracka2020}.
XPS has already been used to characterize the surfaces~\cite{Haasch2000, Walker1998, Haasch2000b, Jaeger2013}
and oxidation~\cite{Milosev1995, Kuznetsov1992} of TiN films,
since even short exposure to air leads to the formation
of oxide and oxynitride overlayer at the surface~\cite{Greczynski2016venting, Greczynski2016selfconsistent}.
As the inelastic electron mean free path in TiO$_2$ is just $\sim$1.5\,nm at 1000\,eV~\cite{Smith2002},
the surface oxide makes it difficult to obtain information about the unoxidized TiN coating itself and thus about the point defects in the film.
For example, it was previously shown that Ti-deficient \TiNx{} is exhibiting some shift of N\,1s
binding energies to lower values~\cite{Walker1998, Delfino1992},
however, the specific explanation, whether this is a feature of the film itself
or just an effect of the oxide layer, was never clarified.
While the straightforward solution to eliminate or greatly reduce the oxygen exposure altogether is to perform the XPS measurements {\it in situ}~\cite{Jaeger2012,Walker1998},
recently, capping approaches were developed to prevent
oxidation of the TiN films even when the samples are exposed to atmosphere~\cite{Greczynski2015, Greczynski2017}.
It was also shown that when the oxidation is small,
angle-resolved XPS can be used to distinguish the signal
coming from below the oxidized region~\cite{Jaeger2012}.
Therefore, it is possible to use XPS to obtain information from the bulk-like region
of the TiN thin film.

In this work, we will examine a series of Al-capped Ti-deficient \TiNx{} thin films with 1.18 $\le x \le$ 1.37 deposited 
using high-power pulsed magnetron sputtering together with density functional theory (DFT) calculations of N\,1s and Ti\,2p BE shifts induced by various point defects.
We will critically assess whether point defects in \TiNx{}, specifically Ti and N vacancies and interstitials, can be accurately detected and quantified using the {\it ab initio}-based XPS correlative method proposed here.

\section{Methodology}

\subsection{Experiment}

\TiNx{} thin films were deposited onto Si (100) substrates at floating potential using reactive high-power pulsed magnetron
sputtering (HPPMS) of a 50\,mm diameter Ti target (99.99 \% purity, EVOCHEM Advanced Materials) with an average power density of
9.4\,W/cm$^2$, 2.5\% duty cycle and 50\,$\mu$s on-time. The depositions were performed in a pure nitrogen atmosphere in order to produce highly Ti-deficient films at a pressure of 1\,Pa and a target-to-substrate distance of 10\,cm for 2 hours. The reason for using pure nitrogen as opposed to more common Ar+N$_2$ mixture was to get as N-rich film as possible and therefore large Ti vacancy concentrations. Some of the films with smaller off-stoichiometry could have been deposited also from the mixed Ar+N$_2$ atmosphere, however, pure nitrogen was used for all samples for consistency. The thin film composition was controlled by varying the substrate temperature systematically, from $T_\mathrm{substrate}$ = 100 to 400\,$^\circ$C in 100\,$^\circ$C increments 
The base pressure at deposition temperature was for all depositions less than $3.6 \times 10^{-5}$\,Pa.
After growth, the as deposited thin films were cooled to room temperature for two hours, while still under vacuum. Subsequently, and before atmosphere exposure, a thin Al capping layer was deposited for 12\,s by direct current magnetron sputtering of a diameter-50\,mm Al target at 3.9\,W/cm$^2$ power density in Ar atmosphere at 0.5\,Pa.
The Al capping layer was deposited at room temperature  since a pronounced island growth was observed at elevated temperatures.
Sputter cleaning behind a closed shutter for at least 2 minutes was performed for both Ti and Al targets before the TiN and Al depositions, respectively.
Al capping was chosen based on the work by Greczynski et al.~\cite{Greczynski2015}, since Al forms a dense oxide scale with only $\sim$4\,nm thickness even after prolonged exposure to air~\cite{vanHarten2009} and is a light element, so XPS signal attenuation is low.
After capping and venting, the samples were immediately transferred to the KRATOS AXIS SUPRA X-ray photoelectron
spectrometer to limit the surface contamination by atmosphere exposure, keeping the total air exposure
time below 10 minutes.

The XPS was equipped with a monochromatic Al\,K$_\alpha$ source and a hemispherical detector.
The spectrometer was calibrated with respect to the BE of Au 4f$_{7/5}$ (83.9\,eV), Ag 3d$_{5/2}$ (368.3\,eV), and Cu 2p$_{3/2}$ (932.7\,eV) signals. 
During spectra acquisition, the base pressure of the system was $< 5.0 \times 10^{-6}$\,Pa.
No charging effects were observed during XPS measurements.
O\,1s, Ti\,2p, N\,1s, C\,1s and Al\,2p core-level spectra were collected with
a pass energy of 20\,eV, a step size of 0.1\,eV, and a dwell time of 1000\,ms in
3 alternating sweeps.
Additional high resolution (20 sweeps, 10\,eV pass energy, 1\,s dwell time, and 0.04\,eV step) and angle-resolved XPS measurements (at 0°, 50°, and 70°, 3 sweeps, 20\,eV pass energy, 1\,s dwell time and 0.05\,eV step) were performed for the N\,1s region.

For the structural analysis with X-ray diffraction (XRD),
a Bruker AXS D8 Discover General Area Detector Diffraction System (GADDS)
was utilized.
The Cu K$_{alpha}$ ($\lambda = 1.5406$\,Å) X-ray source was set to 40\,kV
at a current of 40\,mA and the incident angle was fixed at 15$^\circ$.

Film thicknesses were measured using a FEI Helios Nanolab 660 dual-beam microscope. The region of interest was marked with a Pt protection layer and a trench was milled, employing focused ion beam techniques with Ga ions at 30\,kV. The film thicknesses were determined at an angle of 52\,degrees with tilt correction.

Ion beam analysis has been performed at the Tandem Laboratory of Uppsala University~\cite{Strm2022}. Depth profiles of the elemental composition were obtained by time-of-flight elastic recoil detection analysis (ToF-ERDA) using 36\,MeV $^{127}$I$^{8+}$ primary ions.
The ion beam was directed onto the sample at an angle of 67.5° with respect to the surface normal and time-energy coincidence spectra were obtained at an angle of 45° with respect to the primary beam. The detection system is described in more detail in~\cite{Zhang1999} and a segmented gas detector system was employed~\cite{Strm2016}.
Time-energy coincidence spectra were converted to depth profiles with CONTES~\cite{Janson2004} and all films exhibited homogeneous depth profiles.
Sources of uncertainty for ToF-ERDA are the detection efficiency below unity as well as the specific energy loss of primary ions and recoil species~\cite{toBaben2016}.
Minimization of such uncertainties can be realized by combining different methods~\cite{Moro2018}. Thus, ToF-ERDA was combined with Rutherford backscattering spectrometry (RBS) using a 2\,MeV $^4$He$^{+}$ primary beam and backscattered ions were detected at an angle of 170°.
SIMNRA~\cite{Mayer1999} was used for RBS data analysis.
Based on the combination of both techniques---ToF-ERDA and RBS---the total measurement uncertainty was 2\% relative of the deduced values.
The samples are referenced by their ToF-ERDA/RBS composition throughout the manuscript (e.g., \TiNone{}, \TiNfour{}).

\subsection{Ab initio modelling}

2$\times$2$\times$2 TiN supercells with size of 64 atoms
and a single-point defect were used for the calculations.
The list of point defects considered here consists of N (V$_\mathrm{N}$) and Ti (V$_\mathrm{Ti}$) vacancies,
N and Ti tetrahedral (I$_\mathrm{N}^\mathrm{t}$ and I$_\mathrm{Ti}^\mathrm{t}$), split (10$\overline{1}$)-aligned (I$_\mathrm{N}^{(10\overline{1})}$ and I$_\mathrm{Ti}^{(10\overline{1})}$) and split (111)-aligned (I$_\mathrm{N}^{(111)}$ and I$_\mathrm{Ti}^{(111)}$) N and Ti interstitials~\cite{Tsetseris2007}
as well as of the most stable N Frenkel pair consisting of the
N split (10$\overline{1}$)-aligned interstitial with the two N atoms
equally distant from the vacancy site (FP$_\mathrm{N}^{(10\overline{1})}$)~\cite{Sangiovanni2015}.
Furthermore, 3$\times$3$\times$3 supercell models with a single defect were prepared for V$_\mathrm{N}$ and V$_\mathrm{Ti}$ 
in order to estimate the influence of the supercell size.
The initial models were fully relaxed with respect to atomic positions
and cell size and shape using the Vienna {\it Ab initio} Simulation Package (VASP)~\cite{Kresse1996a, Kresse1996b} DFT~\cite{Hohenberg1964,Kohn1965} code.
Projector augmented wave (PAW) pseudopotentials~\cite{Kresse1999} were used with the plane-wave energy cutoff of 500\,eV,
$k$-point grid of 5$\times$5$\times$5 (3$\times$3$\times$3 for the larger supercells),
and Gaussian smearing method with broadening of 0.1\,eV.

Nudged elastic band method~\cite{Jonsson1998-la} as implemented in VASP was used to estimate the migration barrier between neighboring interstitial sites.
The calculations ran with the same parameters as the structural relaxations above.
We used 5 intermediate images between fully relaxed N in the tetrahedral position and the (111)-aligned split interstitial employing the $2\times2\times2$ supercells.
The images were allowed to be fully structurally relaxed during the NEB calculation. The energy barrier calculated by the NEB method can be used together with the Arhenius equation to estimate transition rates between different interstitial states at a finite temperature.

The N\,1s and Ti\,2p$_{3/2}$ BEs were calculated as the energy difference between the final state,
where a single electron was removed from the specified atom and core level
and placed in the valence band, and the initial ground state, $E_\mathrm{b} = E_\mathrm{final} - E_\mathrm{initial}$.
Self-consistent DFT calculations were used for both the ground and the final state.
The final state always corresponds to the lowest energy state under the constraint
of the specific core electron being removed from the core level and placed into the valence band.
As a result, the method is able to predict the main XPS peaks. However, spectral features
corresponding to additional energy loss events, e.g., the satellite peaks, are not modeled.
We note that while the absolute
BEs do not correspond to the absolute experimental values by few eV, the predicted N\,1s BEs of the TiN main component are at $\sim404.5$\,eV instead of 397.3\,eV~\cite{Haasch2000}, 
the relative BE shifts using the same method for Ti$_{1-\delta}$O$_x$N$_{1-x}$ were previously found to
be in perfect agreement with the experiment~\cite{Ondracka2022}.
The differences in absolute binding energies are to some extent caused by the Coulombic interaction between the core hole and its periodic images. However, the impact of this effect on relative shifts is assumed to be small due to the good electrical conductivity and efficient screening in TiN.
The BE calculations were carried out for every atom in the supercell,
leading to $\sim$64 core hole calculations per defect model and
used the all-electron full-potential
Wien2k code with the linearized augmented plane-wave basis set~\cite{Blaha2020}.
The predicted binding energy shift for atoms in the vicinity of a point defect was calculated as the difference between  their BEs and the average BE of all other atoms of the same type in the structure. As an example, the N 1s BEs of the six equivalent N neighbors around a Ti vacancy were compared to the mean N 1s BE value of the remaining 26 N atoms in the supercell. The reference BEs, labeled TiN, are shown in red in the histograms of Figure~\ref{spec}.
Atomic muffin-tin radii were set to 1.79 and 1.25\,Bohr radius for Ti and N atoms, respectively.
The basis size parameter $R_\mathrm{mt} \cdot K_\mathrm{max}$ of 7.5  
and 4$\times$4$\times$4 $k$-point grid should guarantee
a convergence of better than 0.05\,eV for the calculated BEs 
(while the convergence of the relative energy shifts should be even better).
Perdew--Burke--Ernzerhof (PBE) functional was used for both VASP and Wien2k calculations~\cite{Perdew1996}.
For the purpose of visualization and comparison with experimental
spectra, the discrete calculated binding energies were broadened by Gaussians with $\sigma = 0.3$\,eV.
We note that the switch between two different DFT codes was done for efficiency purposes. VASP offers fast structure relaxation and has a NEB implementation. Wien2k is a slower all-electron code, however, it offers explicit modeling of the core states which benefits the core-hole calculations, e.g., fully relativistic and self-consistent treatment of all core electrons by default.
All DFT calculations are publicly available under
the Creative Commons license in the NOMAD Archive~\cite{Scheidgen2023, Ondracka-TiNdata}.

\section{Results and discussion}

\subsection{Film composition and structure}

\begin{figure}
\begin{center}
\includegraphics[width=0.5\linewidth]{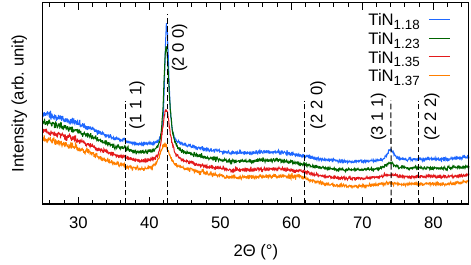}
\caption{\label{fig:XRD} Measured XRD patterns of the \TiNx{} thin films. The dashed lines show expected diffraction peak positions for stoichiometric cubic TiN (PDF Card No.: 00-038-1420).}
\end{center}
\end{figure}

The chemical compositions of all as deposited \TiNx{} films was characterized by ion beam analysis and  are summarized in Table~\ref{tab:composition}.
All measured samples are Ti-deficient with respect to stoichiometric TiN$_{1.0}$ and Ti content
is decreasing with the decreasing deposition temperature.
Figure~\ref{fig:XRD} shows XRD patterns of the deposited films.
All films show a highly textured cubic TiN phase,
however, with the decreasing deposition temperature, the intensity of the diffraction peaks clearly decreases and larger peak broadening is observed, indicating smaller TiN crystallites.
Additionally, the peak shifts to smaller $2\theta$ angles at lower deposition temperatures,
thus corresponding to a lattice parameter increase for films deposited at lower temperatures. 
However, due to very significant texturing in the films and small grain sizes, the evaluation of stress-free lattice
parameter, e.g., by $\sin^2 \Psi$ method, was not possible.
The film thicknesses varied between 120 and 90\,nm (as determined from film cross section measurements shown in Supplementary Information, Figure~SI3) corresponding to deposition rates 1.0--0.7\,nm/min.

\begin{table}
\caption{\label{tab:composition} Normalized film compositions as measured by combination of ToF--ERDA and RBS.
Traces of O (less than 0.7\,at.\% in all samples) and Al from the capping layer (less than 0.2\,at.\% in all samples) were detected as well.
}
\begin{center}
\begin{tabular}{ccccc}
\hline
	$T_\mathrm{dep}$ ($^\circ$C) & Ti (at. \%) & N (at. \%) & $x$ in \TiNx{}\\
\hline\hline
100 & $42.2(12)$ & $57.8(12)$ & 1.37(5)\\
200 & $42.5(12)$ & $57.5(12)$ & 1.35(5)\\
300 & $44.8(11)$ & $55.2(11)$ & 1.23(5)\\
400 & $45.8(11)$ & $54.2(11)$ & 1.18(5)\\
\hline
\end{tabular}
\end{center}
\end{table}

\subsection{Capping layer quality}

\begin{figure}
\begin{center}
\includegraphics[width=0.5\linewidth]{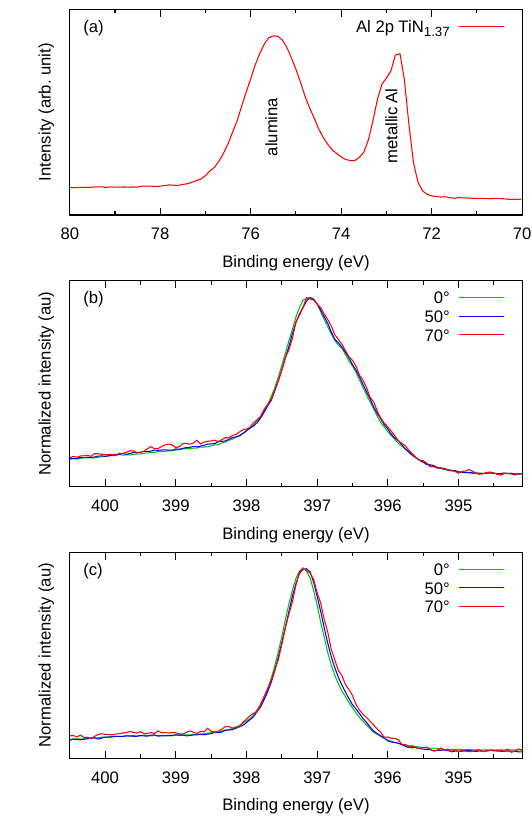}
\caption{\label{capping_check} a) Typical Al\,2p spectrum of the capping layer, similar to what was reported in Ref.~\cite{Greczynski2015}.
N\,1s spectra of b) \TiNone{}, c) \TiNfour{} at different measurement angles.
Measurement angles of 50$^\circ$ and 70$^\circ$ correspond to measurement depths of 64\% and 34\%,
respectively, as compared to measurements at 0$^\circ$ (detector in the direction of the surface normal).}
\end{center}
\end{figure}

A crucial prerequisite before attempting to detect and quantify defects via XPS
is verifying the effectiveness of the deposited capping layer;
otherwise, it is difficult to distinguish between the spectral features
originating in the surface oxidized layer and in the unoxidized film underneath.
An overlayer of TiO$_2$ and/or TiO$_x$N$_y$ is formed, when TiN oxidizes.
Specifically, the TiO$_x$N$_y$ surface component was shown to lie $\sim$1\,eV
below the TiN component in the N\,1s spectra~\cite{Greczynski2016selfconsistent}.
Therefore, a two-step workflow was developed to estimate the effectiveness of the Al capping.
The first requirement is that the capping layer is thick enough.
This was evaluated from the XPS Al\,2p spectra, where the contribution from the surface alumina 
and the remaining metallic Al below can be clearly distinguished.
An optimal amount of Al as a compromise between the signal attenuation in the capping and the oxidation protection
was determined to be when the alumina and metallic Al peak had approximately the same height, as can be seen in Figure~\ref{capping_check}\,a). 

The second important indicator of the Al capping effectiveness is the evolution of the N\,1s peak 
under different XPS measurement angles, shown in Figure~\ref{capping_check}b) and c) for \TiNone{} and \TiNfour{} films, respectively.
The differences between the N\,1s measurements at different angles are negligible
and much smaller than the differences between the different films, which will be discussed later.
Since the XPS signals measured at different angles and thus from different depths are virtually identical, we can conclude that the film is homogeneous at the depths probed by XPS and there are no pronounced effects at the capping-TiN interface influencing BEs.

Importantly, if oxidation of the \TiNx{} film surface would have occurred, it would have resulted in oxygen depth-gradient and, therefore, depth-dependence of the N\,1s signal, which is very sensitive to the oxidation~\cite{Greczynski2016selfconsistent}, should have been identified.
As such N\,1s signal depth-dependence could clearly not be resolved, it is evident that the Al capping applied here is effectively protecting from oxidation due to atmosphere exposure in the depth range probed by XPS.

In conclusion, the Al capping is effectively protecting all here-reported TiN films from oxidation.

\subsection{XPS measurements}

\begin{figure}
\begin{center}
\includegraphics[width=0.5\linewidth]{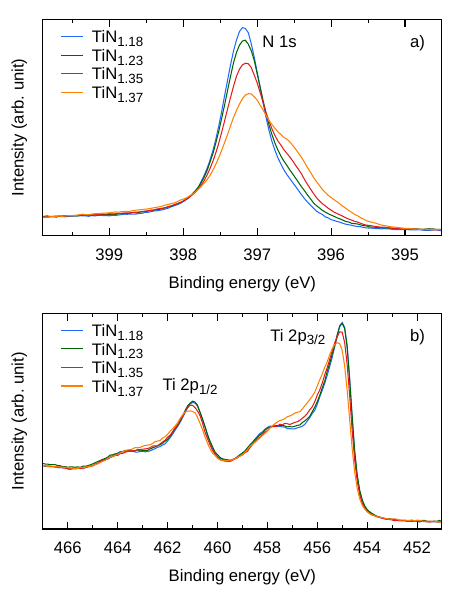}
\caption{\label{N1s} Measured a) N\,1s and b) Ti\,2p XPS spectra. The curves were scaled to identical peak areas and were aligned by the background to highlight the relative changes.}
\end{center}
\end{figure}

XPS measurements of the N\,1s spectra of the deposited films are shown in Figure~\ref{N1s}a).
They exhibit the expected main TiN peak around 397.2\,eV as well as
weak satellite features at higher BEs~\cite{Haasch2000}.
Capped stoichiometric TiN films are expected to only show those two N\,1s components~\cite{Greczynski2016selfconsistent}.
However, with increasing off-stoichiometry, there is a significant shift in the spectral weight from the main
TiN-like peak to lower BEs. 
This is consistent with what was previously reported for Ti vacancies in TiON~\cite{Ondracka2022},
where this signal originates from N atoms with neighboring Ti vacancies.

Measured Ti\,2p XPS spectra are plotted in Figure~\ref{N1s}b).
The spectra exhibit four visible structures,
where the main features around 455\,eV and 461\,eV are attributed to the main
Ti\,2p$_{3/2}$ and Ti\,2p$_{1/2}$ peaks of TiN, respectively,
and the weaker features around 458\, and 464\,eV are satellites~\cite{Haasch2000}.
No other structures are observed.
Ti\,2p BEs exhibit smaller changes with the increasing off-stoichiometry compared to the case of N\,1s.
There is virtually no difference between the \TiNfour{} and \TiNthree{} samples.
As the off-stoichiometry is increased, the main peaks shift to slightly higher BEs and
there is a small increase in spectra intensity between the main and satellite peaks.
Measured XPS overview spectra as well O\,1s, Al\,2p, C\,1s and valence band regions spectra are show in the Supplementary Information, Figures~SI1 and SI2.
We note that the observed oxygen and carbon signals, of varying magnitude from sample to sample, are in excellent agreement with reactions taking place on the Al capping layer: The O\,1s binding energies in the range of 532.2--532.4\,eV is consistent with the formation of Al--O bonds at 532.0–532.4\,eV~\cite{Greczynski2015})while  the C--C/C--H C\,1s peak at 286.1--286.3\,eV conforms to expected BE of adventitious carbon on air-exposed Al at 286.1-–286.2\,eV~\cite{greczynski_reliable_2018}), confirming functionality of the capping layer. Furthermore, there is no evidence for the incorporation of C into the TiN as Ti--C components in the C\,1s spectra $\sim$282\,eV~\cite{cheng_characterization_2007, qin_structure_2018, ou_tribocorrosion_2022} are absent, see Figure SI2b. Therefore, the random variations in the C\,1s and O\,1s intensities are ascribed to the differences in the ambient conditions and air exposure time during venting and sample transport procedures and not to compositional changes in the TiN layer.

Based on the XRD, ion beam analysis and XPS results as well as previously reported DFT calculations~\cite{Zhang2021, Balasubramanian2018, Tsetseris2007},
the following,
preliminary conclusions about the point defects potentially present in the films can be drawn.
There are three possible point defect-based explanations of N excess as measured by the ion beam analysis:
Ti vacancies, N interstitials, and N antisites.
Ti vacancies have the lowest energy of formation of $\sim$3.2\,eV \cite{Zhang2021, Balasubramanian2018}.
N interstitials have energy of formation of 4.6, 4.8 and 5.46\,eV for
the split (10$\overline{1}$)-aligned, split (111)-aligned,
and tetrahedral interstitials~\cite{Tsetseris2007}, respectively. The least energetically favorable defect is the N antisite
with formation energy of $\sim$12.6\,eV~\cite{Zhang2021}.
Therefore, based on thermodynamic considerations, Ti vacancies are the most probable point defects facilitating the observed off-stoichiometry,
due to their low energy of formation.
The shift of the XRD peaks to smaller $2\Theta$ angles
(indicating increasing lattice parameter) for films deposited at lower temperatures with respect to the films deposited at higher temperatures and also
to the powder TiN values, i.e., PDF Card No.: 00-038-1420, may indicate the presence of N  interstitials.
However, since it was not possible to determine the equilibrium lattice parameter,
it is possible that the XRD peak shifts are caused by differences in residual stress states of the films, i.e., larger residual stress is to be expected in the films deposited at higher temperatures.
Therefore, based on the ion beam analysis and XRD data, the presence of N interstitials appear possible but cannot be definitely proven.

In the next step, we critically appraise the above presented hypotheses, i.e., the probable presence of Ti vacancies and/or N interstitials in the as deposited films, based on {\it ab initio} calculations of the influence of specific point defects on the binding energies of the surrounding atoms.

\subsection{Ab initio modeling}

\subsubsection{Core electron binding energies}

\begin{figure}
\begin{center}
\includegraphics[width=0.75\linewidth]{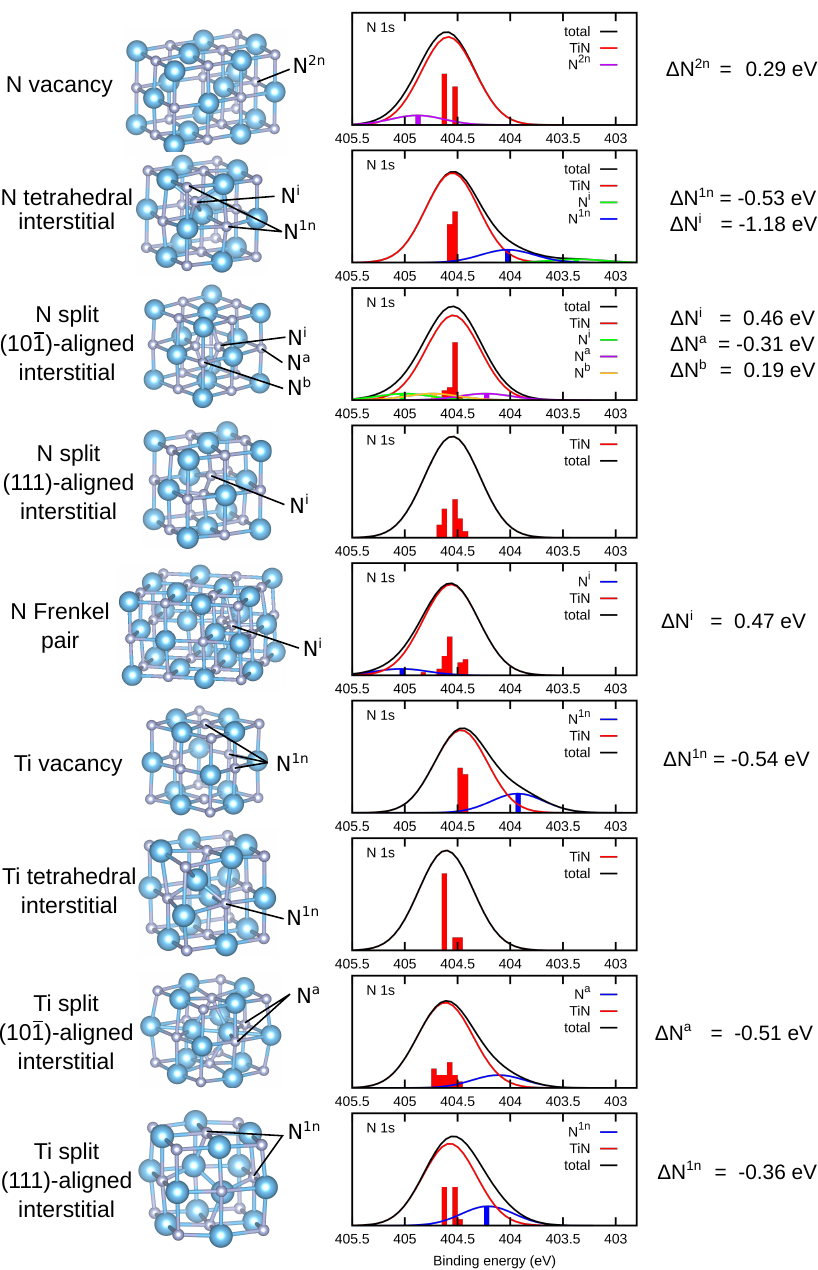}
\caption{\label{spec}First column: atomic structures (visualized by VESTA~\cite{Momma2011}) of the here studied defects with marked atoms exhibiting significant BE shifts.
Second column: histogram and broadened BEs of N\,1s.
Red lines (marked ``TiN'') correspond to all remaining atoms in the supercell, which were not specifically highlighted
due to having none or negligible BE shifts.
Third column: Calculated mean BE shifts with respect to the stoichiometric defect-free TiN (the mean energy of the remaining atoms).
The nomenclature of labeling sites is explained in the corresponding parts of the text.
}
\end{center}
\end{figure}

In the following, we will exclusively discuss the BEs of N\,1s electrons.
The point defect visualizations, BE histograms and broadened BEs, as well
as the mean BE shifts are summarized in Figure~\ref{spec} for all the here studied point defects.
Only atoms with core electron BE shifts ($>0.15$\,eV) are marked in the figures.
We note that this chosen BE shift limit is to large extend arbitrary and the experimental resolvability of specific spectral components depends on multiple conditions, most notably the instrumental resolution, temperature, crystalline quality (structure disorder), signal-to-noise ratio and intensity ratios of the components. In the case of the here studied films, the FWHM of the main N\,1s spectral component is in the range of 0.65--0.8\,eV (as shown in Table~\ref{table:fits} which will be discussed in more detail later) for the measurements at 10\,eV pass energy. As a very crude estimate, two components could be resolved if their BE difference is at least half of this FWHM value.
Therefore, not all of the bellow discussed BE shifts can be resolved by the spectrometer employed in the present work and furthermore,
not all the discussed point defects are expected to be present in the here experimentally studied TiN$_x$ with 1.18 $\le x \le$ 1.37.
Specifically, presence of Ti interstitials or N vare presented in supplementary informationacancies is unlikely, for reasons discussed below.
Nevertheless, the following section is an overview of which point defect-induced BE shifts are relevant for the experimental analysis of point defects in \TiNx{} (both $x \ge 1$ and $x \le 1$) by XPS, assuming in some cases a better instrumentation, e.g., a synchrotron-based XPS.
While our experimental analysis utilizes only predictions of the N\,1s BEs, we have also calculated the Ti\,2p$_{3/2}$ BEs. We propose their potential usefulness for detecting Ti interstitials; however, a rigorous experimental validation is beyond the scope of the present work. Interested reader can find the calculated BEs in the supplementary information as Figure SI5, together with brief discussion.

N vacancy (Figure~\ref{spec}, first row):
The second nearest N$^\mathrm{2n}$ neighbors of the vacancy see a slight increase in BEs by around 0.29\,eV.
This is consistent with previous ELNES calculations, where the influence of N vacancy was also reported to be strongest for the second nearest N neighbors~\cite{Tsujimoto2005}.
Other N atoms considered here show BE changes $< 0.1$\,eV and are hence not resolvable with laboratory XPS systems.
The shift of BEs of the second nearest N vacancy N neighbors is larger with almost 0.3\,eV.
However, the position overlaps with the very broad N\,1s satellite peak observed in experiments~\cite{Haasch2000, Greczynski2016selfconsistent},
which is not possible to be modeled by our DFT approach.
This complicates fitting and introduces further potential uncertainties.

N tetrahedral interstitial (Figure~\ref{spec}, second row):
The interstitial N$^\mathrm{i}$ atom exhibits a reduced BE of 1.18\,eV, while
a BE reduction of 0.53\,eV was predicted for its first N$^\mathrm{1n}$ neighbors.
Both BE shifts of $-0.53$\,eV and $-1.18$\,eV are resolvable by laboratory XPS systems.
All other N atoms are predicted to exhibit unresolvable BE shifts.

N split (10$\overline{1}$)-aligned interstitial (Figure~\ref{spec}, third row):
The two split interstitial N atoms show a BE increase of 0.46\,eV.
The first N$^\mathrm{a}$ neighbors in the direction of the N$_2$ bond exhibit a reduced BE by 0.31\,eV.
The first N$^\mathrm{b}$ neighbors perpendicular to the N$_2$ bond see a BE increase of 0.19\,eV.
The rest of N atoms exhibits a BE shift smaller than 0.1\,eV.
The BE shift of the interstitial N atom of 0.46\,eV is resolvable by laboratory XPS,
however,  the signal is weak as it originates only from two atoms per defect.
Potential characterization would be further complicated by an overlap with the satellite peak,
which is also situated at higher BE
and, therefore, this type of interstitial 
could be difficult to be unambiguously resolved by fitting extra spectral components.

N split (111)-aligned interstitial (Figure~\ref{spec}, fourth row):
There is no N atom with a BE shift $\ge 0.15$\,eV, therefore, this type of N interstitial is expected to only lead to small peak broadening and, thus,
is deemed to be undetectable by XPS.

N Frenkel pair (Figure~\ref{spec}, fifth row):
The calculated Frankel pair defect consists of the N split (10$\overline{1}$)-aligned interstitial 
and a close N vacancy in the direction perpendicular to the N--N bond.
The calculated BE shifts are very close to what would be expected from just adding together the effects of the
before-calculated N vacancy and N split (10$\overline{1}$)-aligned interstitial.
The two interstitial atoms N$^\mathrm{i}$ have their N\,1s BE increased by 0.47\,eV
(as compared to 0.46\,eV for the separate interstitial).

Ti vacancy (Figure~\ref{spec}, sixth row):
N$^\mathrm{1n}$ atoms directly next to the Ti vacancy have reduced BEs by 0.54\,eV
with respect to the rest of the N atoms
and the signal is strong as a single vacancy influences 6 N atoms.
This interpretation is consistent with the previously published predictions for TiON~\cite{Ondracka2022}, where it was
additionally published that the BE shift is approximately additive
with respect to the number of vacancies
and that the effect can be used for Ti vacancy quantification using the XPS.
The BEs shifts of the Ti atoms $\le 0.1$,\,eV, are unresolvable by laboratory XPS.

Ti tetrahedral interstitial (Figure~\ref{spec}, seventh row):
This Ti interstitial induces unresolvable changes in the N\,1s BEs of the surrounding N atoms.

Ti split (10$\overline{1}$)-aligned interstitial (Figure~\ref{spec} eighth row):
The only significantly influenced N atoms are the nearest N$^\mathrm{a}$ neighbors in the (100) plane,
while no other calculated N\,1s binding energies stand out.
The general distribution of the remaining BEs from N atoms is quite broad.
This is expected as Ti introduces a much larger overall displacements than the smaller N interstitials.

Ti interstitial 111 (Figure~\ref{spec}, ninth row):
The first 6 N$^\mathrm{1n}$ neighbors of the Ti interstitial pair have their BEs reduced by 0.36\,eV.
Other N atoms see no significant change.

Therefore, we predict a possibility of detecting point defects which induce
a significant negative BE shift of the N\,1s atoms, such as Ti vacancies or N tetrahedral interstitials
as there is no overlapping peak at lower BEs (assuming the film can be protected from oxidation~\cite{Greczynski2016selfconsistent}).
Nevertheless, it seems difficult to distinguish Ti vacancies and N tetrahedral interstitials from the XPS spectra.
Selected Ti interstitials also induce negative BE shifts in the N\,1s spectra,
however, they are unlikely based on their high formation energy of more than $11$\,eV~\cite{Balasubramanian2018}, compared to 2.4--2.5\,eV for N vacancy~\cite{Tsetseris2007, Balasubramanian2018}. 
Point defects inducing positive BE shifts of the N\,1s core electrons,
like N vacancies and N split (10$\overline{1}$)-aligned interstitials, are predicted to be difficult to detect
due to overlap with the satellite peak located at higher BE with respect to the main TiN N\,1s component.
The effect of possible stress in the films is not taken into account in the first principles BE simulations, they are all done with fully relaxed structures. We have however performed additional calculations, which show that the effect is negligible. Specifically, for the structure with single Ti vacancy with a 0.02\,A lattice parameter difference, the change in both absolute BEs and also the BE shifts is less than 0.01\,eV, which we deem undetectable by table-top XPS.
 
\subsubsection{Ab initio lattice parameters}

\begin{table}
\caption{\label{tab:latpar}Calculated directionally-averaged lattice parameter $a$, corresponding to the specific defect concentration of 1 per 64 atoms (in a reference perfect lattice) and its difference from the value for stoichiometric defect-free TiN (labeled as ``none'').}
\begin{center}
\begin{tabular}{lcc}
\hline\hline
Defect type & $a$ (Å) & $\Delta a$ (Å)\\
\hline
none & 4.239 & 0\\
V$_\mathrm{N}$ & 4.239 & $<-0.001$\\
V$_\mathrm{Ti}$ & 4.230 & $-0.009$\\
I$_\mathrm{N}^{(10\overline{1})}$ & 4.267 & 0.028\\
I$_\mathrm{N}^{(111)}$ & 4.268 & 0.029\\
I$_\mathrm{N}^\mathrm{t}$ & 4.267 & 0.028\\
FP$_\mathrm{N}^{(10\overline{1})}$ & 4.267 & 0.028\\
I$_\mathrm{Ti}^{(10\overline{1})}$ & 4.296 & 0.057\\
I$_\mathrm{Ti}^{(111)}$ & 4.294 & 0.055\\
I$_\mathrm{Ti}^\mathrm{t}$ & 4.296 & 0.057\\
\hline\hline
\end{tabular}
\end{center}
\end{table}

It was already shown that the lattice parameter can be one of the factors signaling the presence of point defects~\cite{Zhang2021}.
Table~\ref{tab:latpar} shows the predicted impact of the studied point defects on the lattice parameter,
which corresponds to the concentration of one defect per 64 atoms (in the reference perfect lattice).
In general, at the studied defect concentration, N vacancies have no significant effect on the lattice parameter,
while Ti vacancies reduce the lattice parameter by $\sim$0.01\,Å at the simulated vacancy concentration.
N interstitial increases the lattice parameter by around 0.03\,Å independent of the interstitial type,
while an increase in lattice parameter of 0.06\,Å is predicted for Ti interstitial
and again does not depend on its type.
N Frenkel pairs influence the lattice parameter in a similar way as N interstitials alone.

\subsection{Ti vacancy quantification}

{
Based on DFT calculations, the observed changes in the N\,1s spectra with increasing off-stoichiometry
can be explained by Ti vacancies, N tetrahedral interstitials or Ti interstitials.
The formation of Ti interstitials can be ruled out, as the Ti interstitials have a high formation energies and are furthermore not favored by the N-rich film stoichiometry. 
On the contrary, the presence of both Ti vacancies and N tetrahedral interstitials is consistent with the film composition measured by the ion beam analysis (Table~\ref{tab:composition}) and XRD (Figure~\ref{fig:XRD}) results.
Only the tetrahedral N interstitials are predicted to be detectable in the XPS spectra.
The split N interstitials are more stable than the tetrahedral ones~\cite{Tsetseris2007}, 
however, as HPPMS deposition is a highly energetic process, 
creating N tetrahedral interstitials cannot be ruled out.
Because the N tetrahedral interstitials have a fingerprint that is indistinguishable
from the fingerprint of the Ti vacancies, while the split N interstitials do not have a signal overlapping with the signal of Ti vacancies,
it is important to estimate the respective N interstitial populations.

Results from our NEB calculations show,
that the energy barrier $E_\mathrm{a}$ for the transition of N interstitial
from the tetrahedral position to the geometrically closest (111) aligned
split interstitial is only $\sim$0.35\,eV.
The transition frequency can be calculated as 
\begin{equation}
k = A e^{-\frac{E\mathrm{a}}{k_\mathrm{B}T}} \,.
\end{equation}
where $A$ is the attempt frequency, $k_\mathrm{B}$ is the Boltzmann constant and $T$ is the temperature.
Using the before-calculated activation energy
together with a very crude estimate of the attempt frequency at 5\,THz,
based on the position of the first maximum in TiN phonon density of states~\cite{Gupta2014},
leads to transition frequency of 6.6\,MHz at room temperature and 94\,MHz at $100\,^\circ\mathrm{C}$.
Hence, it is predicted that a large population 
of tetrahedral N interstitials is unlikely, even at room temperature and, therefore, the vast majority of the N interstitials
are predicted to be the split ones, which have no distinct signal in the XPS N\,1s spectra
and are predicted to only lead to peak broadening of the main N\,1s component
(or would overlap with the satellite).
Therefore, for the following analyses, it is reasonable to assume that all the N\,1s XPS signal at the lower BE
peaks originates from the presence of Ti vacancies.

For the Ti vacancy quantification in \TiNx{}, we adapt a method published
previously by us for the analysis of Ti vacancies in TiON thin films~\cite{Ondracka2022}.
A similar fitting procedure was used to fit the N\,1s peak components, with the most notable
difference is that there was no sign of the N$_2$ peak, so it was not included during the fitting.
Four peaks were used for the fits, corresponding to the main TiN component
(Ti atoms with 6 N neighbors and 0 vacancies, denoted 0V),
two components at lower BEs corresponding to Ti atoms with 1 and 2 vacancies
(denoted 1V and 2V) in the first coordination shell and a satellite peak.
Approximate Voigt (Gaussian/Lorentzian) peakshapes in a product form~\cite{Evans1991}
\begin{equation}
    GL(p_i) = \mathrm{exp}\left( -4\ln{2} (1-p_i) \frac{(E-E_i)^2}{B_i^2}\right) /(1+4p_i\frac{(x-E)^2}{B^2})  \, ,
\end{equation}
where $E$ is energy, $E_i$ is peak position, $B_i$ is peak broadening parameter (its value is close to full width at half the maximum intensity), and $p_i$ stands for fraction of the Lorentzian part in \% of a peak $i$ were employed.
The GL(70) lineshape was used for the main TiN-like (0V) peak, while GL(30) line shapes were utilized for the 1V and 2V peaks and GL(0) for the satellite peak.
Peak intensities, positions, and broadening were free-fitting parameters,
however, only a single shared broadening parameter was fitted for the 1V and 2V peaks to reduce fit ambiguity.
A Tougaard-like background shape was utilized during the fitting.
The fits of the N\,1s regions of all samples are shown in Figure~\ref{fits}, while the fit parameters
are summarized in Table~\ref{table:fits}.
Using this four-peak model, it was possible to achieve near-perfect fits for all the studied samples.
Additionally, the fit parameters are very stable between the fits
and the fitted peak positions are in good agreement with the predicted shifts.
Specifically, the positions of the three main peaks are in the range of 397.12--397.19\,eV, 396.49--396.52\,eV,
and 395.91--396.07\,eV for the 0V, 1V, and 2V components, respectively.
The 1V and 2V components exhibit negative shifts of 0.62--0.67\,eV and 1.12--1.21\,eV, respectively. These ranges reflect the variability among the four fits of different samples, relative to 0V (TiN) component.
This is in excellent agreement with the here predicted value of $-0.54$\,eV for the shift between 0V and 1V peak
as well as the predicted and experimentally observed shifts
between corresponding components in TiON~\cite{Ondracka2022}.

The fitting was followed by a simplified version of the analysis of the relative peak ratios
from~\cite{Ondracka2022}, enabled by the fact that there is no oxidation of the TiN layer due to the Al capping,
and, therefore, no need for corrections compensating the TiN oxidation.
The simplified analysis comes at the expense of more complex deposition procedure and can limit the applicability of this method in some cases, e.g., if the Al deposition step can not be easily added to the existing deposition workflow.
Vacancy concentration can be directly quantified from the three N\,1s main components (0V, 1V, 2V) ratios.
Assuming a cubic TiN monocrystal, i.e., neglecting the grain boundaries, with random vacancy distribution,
the probability of finding an N atom with $6-n$ Ti first neighbors,
or equivalently with $n$ neighboring Ti vacancies, is~\cite{Ondracka2022}
\begin{equation}
P^{\NV{n}}(\delta) = {6 \choose {n}} \delta^{n} (1-\delta)^{6-n} \mathrm{,}
\label{prob}
\end{equation}
where $\delta$ is the vacancy concentration on the metal sublattice.

During the vacancy quantification, we search for a $\delta$ which minimizes the squared differences
between the experimental (fitted) $\NV{n}$ components and the theoretical ones,
where both the experimental and theoretical functions were previously renormalized to sum to 1.
Only the first three components are considered, i.e., corresponding to the N atoms with 0, 1, and 2 Ti vacancies in the first coordination shell.
The normalization is needed for the experimental components, as opposed to using directly 
the fractions from Table~\ref{table:fits}, since the fit also includes the satellite peak.
For the theoretical $P^{\NV{n}}(\delta)$ probabilities, the normalization is equivalent to neglecting
the components corresponding to the N atoms with more
than two vacancies in the first coordination shell.
There is no experimental evidence for a corresponding signal
in our data and the probability of finding such
configuration at small $\delta$ is very low; therefore, this is a reasonable simplification.
This approach is consistent with that employed in Ref.~\cite{Ondracka2022},
but it is significantly more robust due to the capping layer,
eliminating the need for additional parameters to account for oxidation,
which in turn reduces uncertainty.

\begin{table}
\caption{\label{table:fits} Fitted parameters for the fits shown in Figure~\ref{fits}.
$f_i$, $E_i$ and $B_i$ correspond to peak fraction (of the total fitted area),
peak position and peak broadening parameter for the $i$ component.}
\setlength\tabcolsep{4pt}
\begin{center}
\begin{tabular}{lccc}
\hline\hline
sample  & $f_\mathrm{TiN}$ & $E_\mathrm{TiN}$\,(eV) & $B_\mathrm{TiN}$\,(eV) \\
TiN$_{1.37}$ & 0.485(4) & 397.123(2) & 0.785(4) \\
TiN$_{1.35}$ & 0.577(3) & 397.159(1) & 0.714(2) \\
TiN$_{1.23}$ & 0.645(3) & 397.1754(8) & 0.686(2) \\
TiN$_{1.18}$ & 0.672(3) & 397.1867(7) & 0.665(2) \\
\hline
sample  & $f_\mathrm{1v}$ & $E_\mathrm{1v}$\,(eV) & $B_\mathrm{1v}$\,(eV) \\
TiN$_{1.37}$ & 0.147(3) & 396.495(4) & 0.65(1) \\
TiN$_{1.35}$ & 0.106(2) & 396.513(4) & 0.6(1) \\
TiN$_{1.23}$ & 0.064(2) & 396.51(5) & 0.54(1) \\
TiN$_{1.18}$ & 0.044(2) & 396.504(7) & 0.48(2) \\
\hline
sample  & $f_\mathrm{2v}$ & $E_\mathrm{2v}$\,(eV) & $B_\mathrm{2v}$\,(eV) \\
TiN$_{1.37}$ & 0.046(1) & 395.916(8) & 0.65(1) \\
TiN$_{1.35}$ & 0.0237(9) & 395.95(1) & 0.6(1) \\
TiN$_{1.23}$ & 0.016(1) & 396.01(2) & 0.54(1) \\
TiN$_{1.18}$ & 0.012(1) & 396.07(2) & 0.48(2) \\
\hline
sample  & $f_\mathrm{satel}$ & $E_\mathrm{satel}$\,(eV) & $B_\mathrm{satel}$\,(eV) \\
TiN$_{1.37}$ & 0.322(3) & 398.07(2) & 3.42(3) \\
TiN$_{1.35}$ & 0.293(3) & 398.22(2) & 3.46(3) \\
TiN$_{1.23}$ & 0.274(3) & 398.34(2) & 3.48(4) \\
TiN$_{1.18}$ & 0.272(3) & 398.38(2) & 3.46(4) \\
\hline\hline
\end{tabular}
\end{center}
\end{table}

\begin{figure}
\begin{center}
\includegraphics[width=0.5\linewidth]{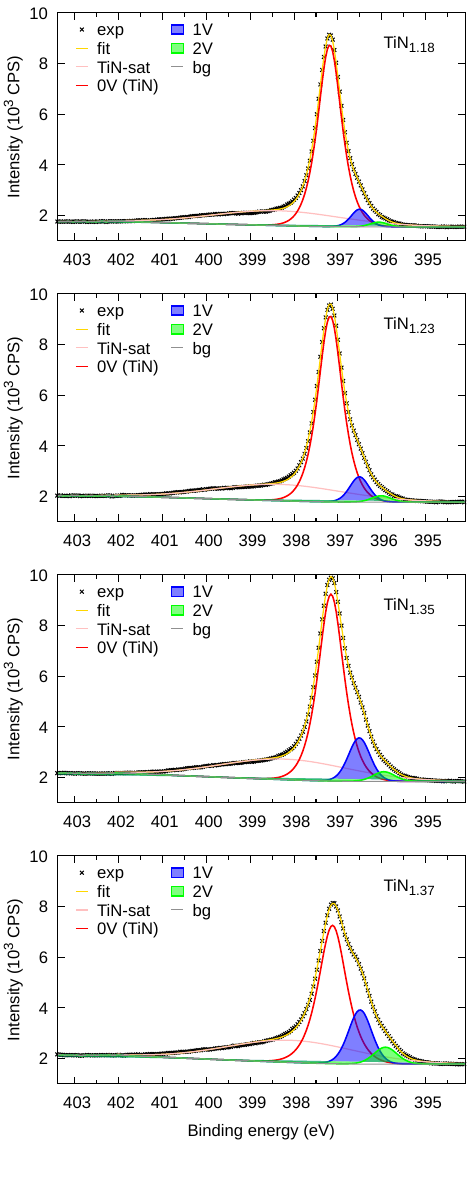}
\caption{\label{fits} Fits of XPS N\,1s spectra of all samples using the four-peak model discussed in the text.}
\end{center}
\end{figure}

The theoretical $P^{\NV{n}}_\mathrm{norm}(\delta)$ are shown in Figure~\ref{delta}
together with the fitted ratios for the four here studied samples.
The $x$ axis positions of the experimental points correspond to the fitted $\delta$ using the aforementioned procedure and is thus the determined Ti vacancy concentration on the metal sublattice. The determined vacancy concentrations  are $1.2(2)$, $1.9(4)$, $3.4(5)$, and $6(1)$ percent of Ti vacancies on the metal sublattice for the \TiNfour{}, \TiNthree{}, \TiNtwo{} and \TiNone{}, respectively. The vacancy concentrations are lower than would be expected based on the ERDA/RBS film stoichiometry under the assumption of Ti vacancies only, and the most likely explanation for this discrepancy is presence of N interstitials in the films. The insensitivity of the XPS-based point defect detection method to the presence of N interstitial therefore can be considered as a strength of the method when determining the Ti vacancies, as opposed to possible estimates from the stoichiometry or a stress-free lattice parameter, since the stoichiometry or lattice parameter measurements are always influenced by the sum of effects from both N interstitials and Ti vacancies with no option to distinguish the individual contributions.
It most be noted that in some cases (most notably with \TiNone{}), that the experimentally determined 0V, 1V and 2V component ratios are not a perfect fit for the theoretically-modeled component ratios. Specifically, the experimental 2V component ratio would better correspond to higher $\delta$ value, while the experimental 0V and 1V ratios would suggest a lower $\delta$ value. As noted before, our model in Eq.~(\ref{prob}) assumes a random vacancy distribution. If the vacancy distribution is not completely random, some disagreement between the theoretical and experimental component ratios would be expected. It would be therefore theoretically possible to detect not only the vacancy concentration but also provide some information about their distribution, however, such a theoretical model is beyond the scope of this work.

\begin{figure}
\begin{center}
\includegraphics[width=0.5\linewidth]{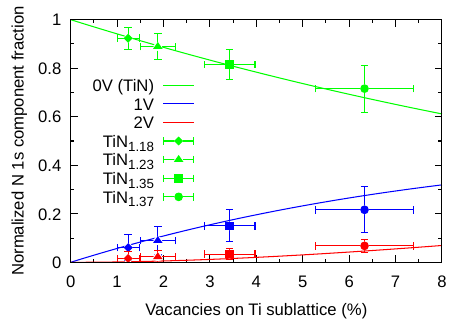}
\caption{\label{delta} Continuous lines show theoretical $P^{\NV{0}}_\mathrm{norm}(\delta)$ (0 vacancies, 0V), $P^{\NV{1}}_\mathrm{norm}(\delta)$ (1 vacancy, 1V), and $P^{\NV{2}}_\mathrm{norm}(\delta)$ (2 vacancies, 2V) normalized functions corresponding to normalized probabilities to find
N atom with 0, 1, and 2 Ti vacancies in the first coordination shell, assuming
a random vacancy distribution.
Points correspond to the fitted normalized peak ratios for the experimental \TiNx{} samples.
Their $x$ axis position is the determined vacancy concentration $\delta$ in the specific sample corresponding to the minimal weighted (with weight based on the uncertainties from the N\,1s fits) least square difference between their values and the corresponding normalized probabilities
and $x$ axis error bars visualize the $\delta$ uncertainty from this fit.
The $y$ error bars were multiplied by a factor of 25 for visualization purposes.
}
\end{center}
\end{figure}

We note that the N\,1s fits and thus also the $\delta$ results are sensitive to the peak shapes used during the fitting.
The specific combination of the used lineshapes described above was
selected since it leads to fits with the lowest residual mean square errors.
Furthermore, the notion of using different peak shapes for the main TiN-like 0V peak
is based already on previous work~\cite{Greczynski2016selfconsistent}
where this peak was also fitted with a more Lorentzian peak shape.
However, it is not immediately clear why a different lineshape should be needed for the main
TiN-like component and 1V and 2V components.
One possible explanation for the need for a different lineshape for
the TiN component is the presence of interstitials.
We have previously discussed that the presence of split N interstitials is likely and they would overlap with 
the main 0V peak and could contribute to a different broadening character.
To estimate the uncertainty related to the used peak shapes we have also run several different fits,
e.g., with GL(30) or GL(70) for all three main peaks and several other variants (GL(30) also for the satellite peak, GL(90) for the TiN-like peak, etc.).
The average difference between the resulting $\delta$ values of different fits is 0.0025,
with the maximum $\delta$ value difference between different fits was less than 0.01.
Thus, this uncertainty due to peak shape selection is smaller than
the $\delta$ uncertainty from the fit itself which is shown in Figure~\ref{delta}.
Another possible source of uncertainty,
the background shape choice (e.g., linear, Shirley, Tougaard) does not significantly influence the results 
due to a minimal increase of the background over the N\,1s peak
and thus its influence on the $\delta$ results is negligible. 

\section{Conclusions}

Based on {\it ab initio} predictions, it is shown that several point defects in TiN, specifically Ti vacancies and N tetrahedral interstitials induce N\,1s core electron binding energy shifts in the first N  neighbors,
with a magnitude large enough to be detectable with standard laboratory X-ray photoelectron spectroscopy.
The here presented calculations predict that significant N\,1s BE shifts are
associated with N tetrahedral interstitials and Ti vacancies, with N\,1s shifts of -0.54\,eV for first N neighbors of the Ti vacancy and N\,1s shifts of -0.54\,eV for first N neighbors of the tetrahedral N interstitial.
However, while it is not possible to distinguish the Ti vacancies from
N tetrahedral interstitials by the evaluation of N\,1s binding energy shifts,
we have also shown that the N tetrahedral interstitial is not stable at the room temperature.
It quickly transform to the lower energy split interstitials and the population of tetrahedral N interstitials in TiN is negligible as a result.
Therefore, we have shown that the N\,1s signal at lower binding energies can be unambiguously attributed to Ti vacancies, Ti vacancies can be unambiguously detected in the XPS measurements of Al-capped TiN
and the recently published XPS Ti vacancy quantification method for \TiON{} thin films can be adapted also for TiN films. By showing that the N\,1s BE shift in the vicinity of a metal vacancy is independent of the presence of O in the film, we also demonstrate the potential for application to other NaCl-structured nitrides.
The proposed method was successfully experimentally verified by XPS measurements of Al-capped \TiNx{} thin films with 1.18(5) $\le x \le$ 1.37(5), where the Ti vacancy concentration on the metal sublattice ranged from $1.2(2)$\,\% to $6(1)$\,\%, respectively.
Importantly, our application of the Al capping simplifies significantly the previously proposed vacancy quantification methodology, leads to more robust results and we have also presented a rigorous capping quality validation method based on the angle-resolved XPS.
Hence, N rich TiN accommodates off-stoichiometry by both, formation of Ti vacancies as well as N split interstitials. Our results highlight that the {\it ab initio}-guided XPS-based point defect characterization method is suitable for the detection and quantification of several point defects in Al-capped \TiNx{}.

\section*{CRediT authorship contribution statement}
Pavel Ondračka: Conceptualization, Formal analysis, Investigation, Methodology,
Writing - original draft, review \& editing, Visualization.
Pauline Kümmerl: Investigation, Writing - original draft, review \& editing,
Marcus Hans: Investigation, Formal analysis, Writing - review \& editing.
Stanislav Mráz: Investigation, Writing - review \& editing.
Daniel Primetzhofer: Investigation, Formal analysis, Writing - review \& editing.
David Holec: Conceptualization, Investigation, Supervision, Writing - review \& editing.
Petr Vašina: Supervision, Writing - review \& editing, Funding acquisition.
Jochen M. Schneider: Conceptualization, Supervision, Project administration, Writing - review \& editing, Funding acquisition.

\section*{Acknowledgments}
This research was funded by German Research Foundation (DFG, SFB-TR 87/3) ''Pulsed high power plasmas for the synthesis of nanostructured functional layer''.
This work was supported by the Ministry of Education, Youth and Sports of the Czech Republic through the project LM2023039.
The authors also gratefully acknowledge the computing time granted by the JARA Vergabegremium
and provided on the JARA Partition part of the supercomputer CLAIX at RWTH Aachen University
(project JARA0151) and computational resources provided by the e-INFRA CZ project (ID:90254),
supported by the Ministry of Education, Youth and Sports of the Czech Republic.
The NEB calculations have been achieved using the Vienna Scientific Cluster (VSC).
Accelerator operation at Uppsala University has been supported by the Swedish research council VR-RFI (\#2019\_00191).

\section*{Data availability}
{\it Ab initio} simulations are available under the CC BY 4.0 licence in the NOMAD archive~\cite{Ondracka-TiNdata}. Experimental data will be made available on request.

\bibliographystyle{elsarticle-num} 
\bibliography{bib-db}

\end{document}


\maketitle

\section{XPS}

\begin{figure}[h!]
    \centering
    \includegraphics[width=\linewidth]{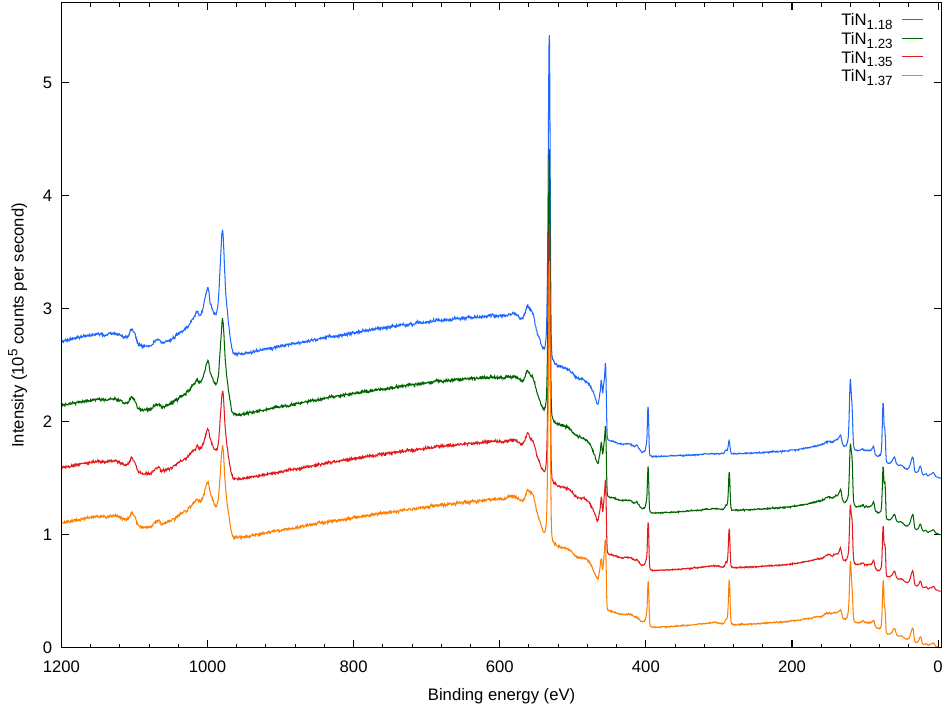}
    \caption{Overview XPS spectra of all films.}
\end{figure}

\begin{figure}[h!]
    \centering
    \includegraphics[width=\linewidth]{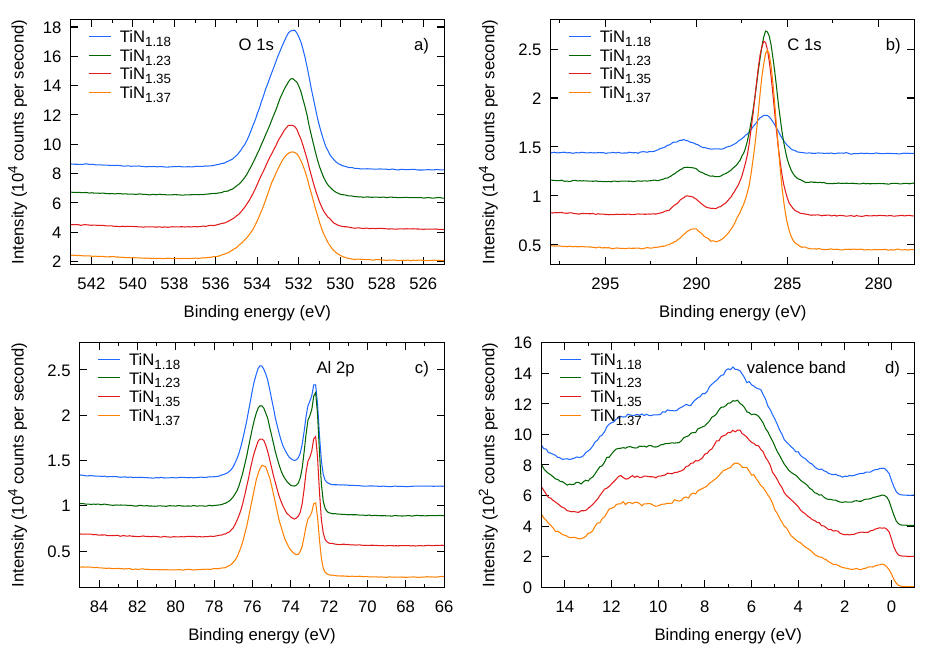}
    \caption{XPS spectra of a) O\,1s, b) C\,1s, c) Al\,2p and d) valence band regions.}
\end{figure}

\clearpage

\section{Film cross sections}

\begin{figure}[h!]
    \centering
    \includegraphics[width=\linewidth]{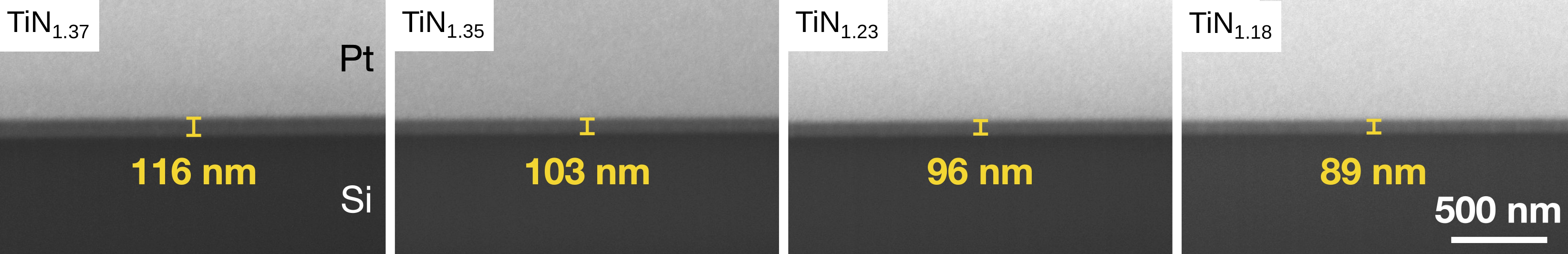}
    \caption{Film cross section measurements.}
\end{figure}

\section{RBS/ERDA}

\begin{figure}[h!]
    \centering
    \includegraphics[width=0.6\linewidth]{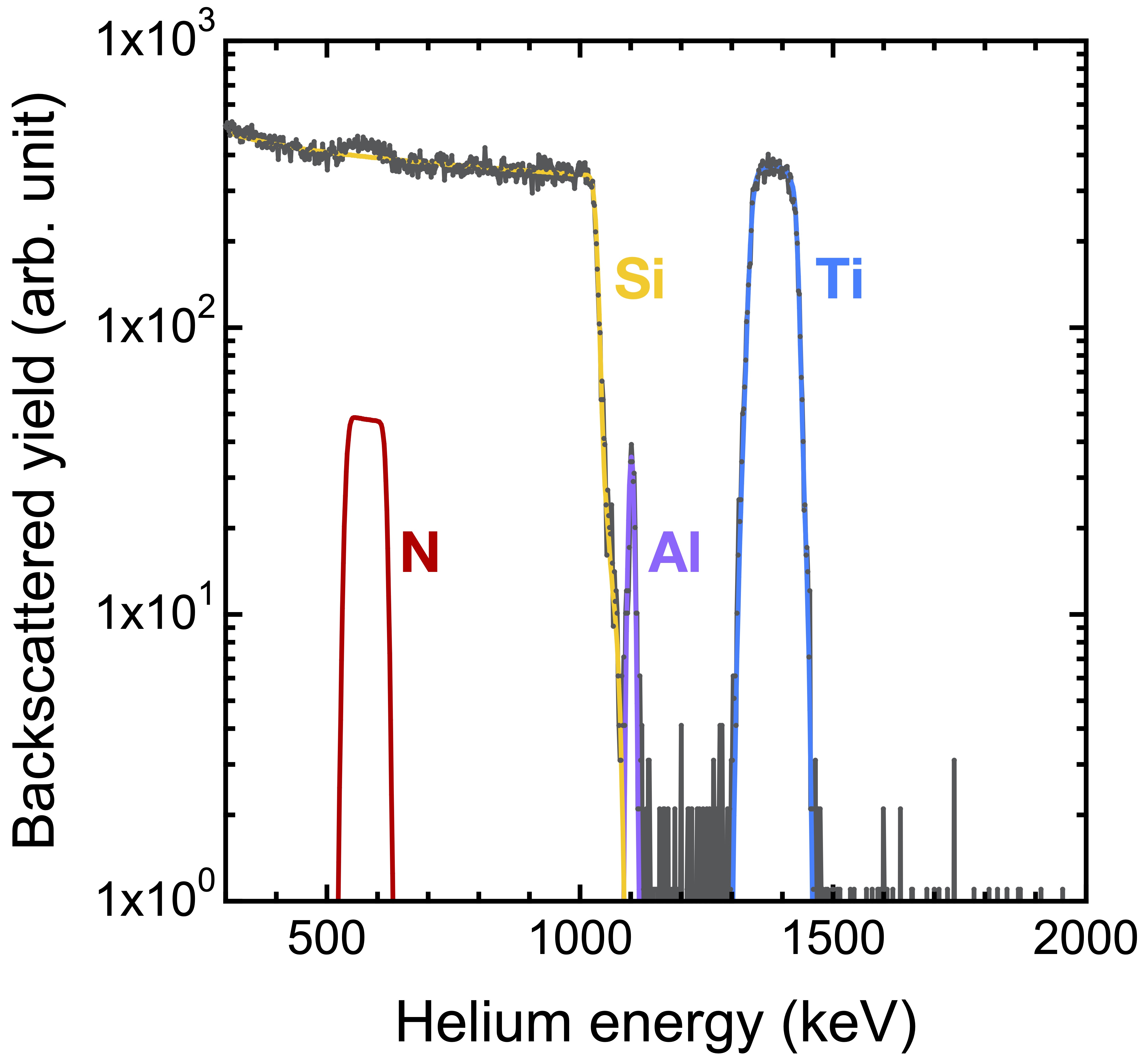}
    \caption{Example Rutherford backscattering spectrum of the TiN$_{1.37}$ thin film. Experimental data points are shown together with simulations of Ti (blue), N (red), Si (gold; from substrate) and Al (purple; from capping layer).}
\end{figure}

\newpage
\section{Ti\,2p$_{3/2}$ simulations}

\begin{figure}[h!]
\begin{center}
\includegraphics[width=0.85\linewidth]{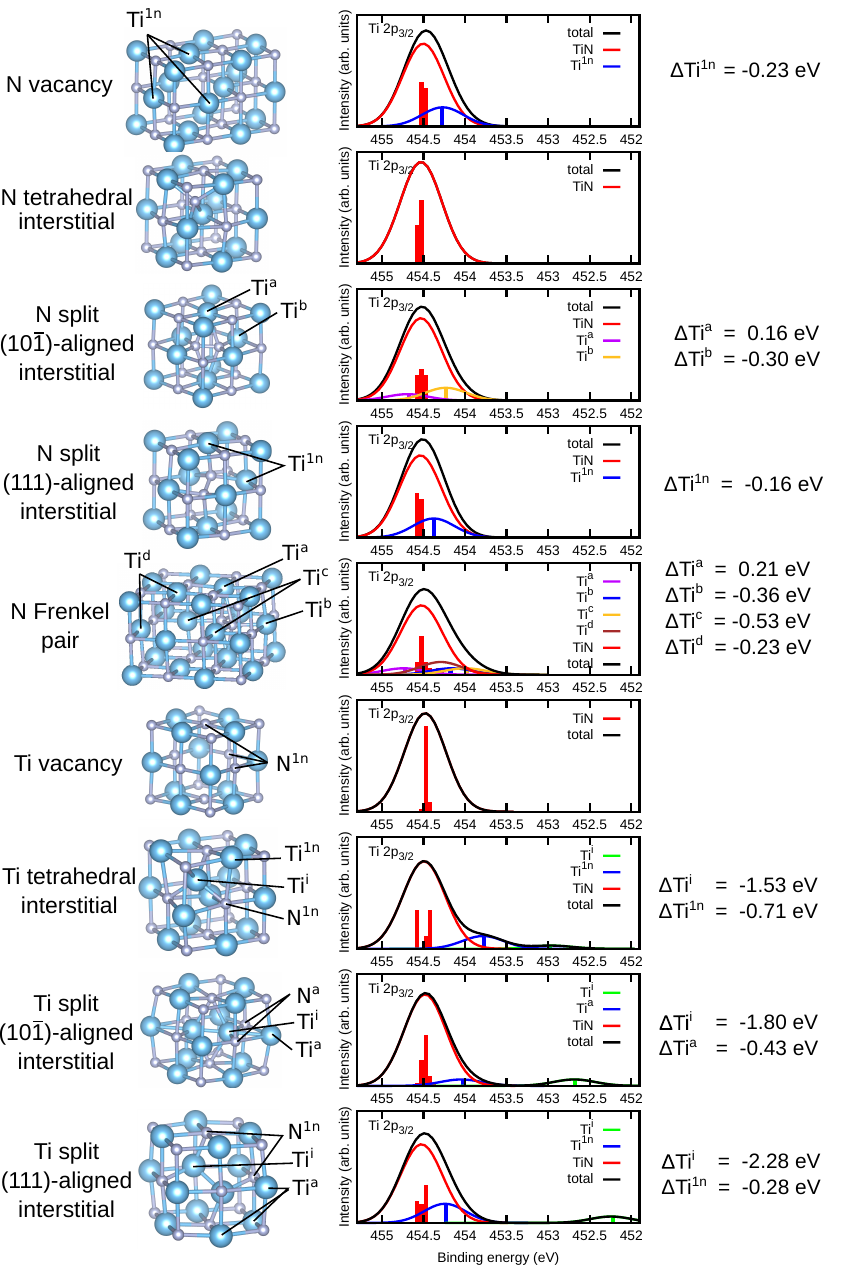}
\caption{\label{spec}First column: atomic structures (visualized by VESTA) of the here studied defects with marked atoms exhibiting significant BE shifts.
Second column: histogram and broadened BEs of Ti\,2p$_{3/2}$.
Red lines (marked ``TiN'') correspond to all remaining atoms in the supercell, which were not specifically highlighted due to having none or negligible BE shifts.
Third column: Calculated mean BE shifts with respect to the stoichiometric defect-free TiN (the mean energy of the remaining atoms).
}
\end{center}
\end{figure}

Ti2p\,2p$_{3/2}$ BE calculations predictions shown in Figure SI5 highlight that all Ti interstitials
exhibit a distinct signal in the Ti\,2p spectra.
Although the presence of Ti interstitials is quite unlikely due to having very high formation energy of more than $11$\,eV~\cite{Balasubramanian2018}, compared to 2.4\,eV for N vacancy~\cite{Tsetseris2007},
so N vacancies are more likely in the case of Ti-rich stoichiometry,
there are no other overlapping signals at lower BEs with respect to the Ti\,2p TiN signal and the predicted Ti\,2p BE shift is $>$1.5\,eV for all Ti interstitial types.
Therefore, even very small Ti interstitial concentrations should be detectable by laboratory XPS.
While we are not aware of any experimental reports of Ti interstitials in TiN, metal interstitials (in Frenkel pairs) were recently suggested in TiAlN~\cite{Holzapfel2022}, therefore here we highlight the XPS as a possible method to either prove or disprove metal interstitial presence in NaCl-type nitrides.
In general, the fitting of Ti\,2p peak is challenging due to the overlap between the Ti\,2p$_{1/2}$ and Ti\,2p$_{3/2}$ peaks
and the presence of satellite peaks which are much more pronounced than in the N\,1s signal.
The choice of background fitting function
and also the choice of a proper asymmetric line shapes is crucial~\cite{Greczynski2016selfconsistent}.
Based on the discussion above, it is predicted that only the Ti interstitials
are readily identifiable from the Ti\,2p spectra.

\bibliographystyle{elsarticle-num} 
\bibliography{bib-db}